\documentclass[english,showpacs,secnumarabic,showkeys]{revtex4}
\usepackage[T1]{fontenc}
\usepackage[latin9]{inputenc}
\usepackage{amsmath}
\usepackage{esint}



\usepackage{babel}

\begin{document}

\title{A Dual Four Dimensional Superstring}

\author{B. B. Deo}

\affiliation{Department of Physics, Utkal University, Bhubaneswar-751004, India}

\email{bdeo@iopb.res.in}

\author{P. K. Jena}

\affiliation{72, Dharma Vihar, Khandagiri, Bhubaneswar-751030, India}

\email{prasantajena@yahoo.com}

\date{\today}



\begin{abstract}
The 26 dimensional bosonic string, first suggested by Nambu and Goto,
is reduced to a four dimensional superstring by using two species
of 6 and 5 Majorana fermions as proposed by Deo. These two species
of fermions differ in their `neutrino-like' phase, and are vectors
in the bosonic representation $SO(d-1,1)$. Using Polchinski's equivalence
between operators and states, we can write the Virasoro generators
for 4 dimensional string theory. The theory is shown to give the same
results as given by other superstrings and also reveals the well known
aspects of four dimensional string theory. The bosons and the fermions
are found to be the basis for constructing this string theory which
includes gravity and exhibits strong-weak coupling duality as well
as the usual electric-magnetic duality. This formalism is used to
calculate the metric tensor as well as the entropy-area relation for
a black hole.
\end{abstract}
\maketitle

\section{Introduction}

Nambu\citet{key-1} and Goto\citet{key-2} proposed a classical relativistic
string theory which turned out to be valid only in 26 dimensions.
It was raised to the quantum level by Goddard, Rebbi, Thorn\citet{key-3},
Goldstone\citet{key-4} and Mandelstam \citet{key-5}. The suggestion
of Scherk and Schwarz\citet{key-6,key-7} that the string theory carries
the quantum numbers for all the four fundamental interactions including
gravity did not make much progress till 1984. Then Green and Schwarz\citet{key-8}
formulated the superstring theory in ten dimensions. The heterotic
string theory of Gross, Harvey, Martinec and Rohm\citet{key-9} was
found to be the first candidate to explain the physical interactions.
Casher, Englert, Nicolai and Taormina\citet{key-10} have proposed
that a 26 dimensional bosonic string contains the ten dimensional
superstring, the two $N=1$ superstrings and the two \textbf{$N=2$}
superstrings. However, it has been a long standing problem to come
down satisfactorily to the four dimensional physical world. Kaku\citet{key11}
and Green, Schwarz and Witten\citet{key-12}, in their books, have
rightly spelt out that `No one really knows how to break the \textit{ten
dimensional theory down to four}'.

The simplest way appears to be to descend directly from the 26-dimensional
bosonic string to the 4-dimensional superstring by using the Mandelstam
equivalence between fermions and bosons in an anomaly free string
theory. The bosons are four in nature. The fermions belong to SO(3,1)
bosonic representation and are divided into two groups. One group
has 24 neutrino-like spinors placed right handedly in six ways and
the other 20 of the similar spinors are placed left handedly in five
ways. Thus the total number of fermions is $4\times6=24$ and $4\times5=20$,
which have opposite handedness. The total number of bosons is equal
to four. These can be taken as the basic objects for constructing
a four dimensional string theory.

One of the present authors(BBD) has shown, in 2003, how to construct
a correct superstring in four dimensions\citet{key-13} from the original
26 dimensional theory. This formulation has been used to find\citet{key-14,key-15,key-16}
most of the properties which would be found by using the ten dimensional
superstrings. There are, of course, many interesting aspects of physics
which cannot be directly found by this method. It was seen\citet{key-17,key-18}
that the equations of motion derived from the low energy effective
action in four dimensional string theory are invariant under the electric-magnetic
duality transformation\citet{key-19} that interchanges the electric
and magnetic fields, and at the same time interchanges the strong
and weak coupling limits. This will be pursued further in four dimensional
theory.

This theory has the correct dimensionality of 26 and in the second
of these, one expects 52 `dimensions' for the 52 types of fields and
field components. Everywhere the ordinary physical dimensions will
be four. There will be sixteen such and four more will be like connecting
factors.

This four dimensional string theory is used successfully in black
hole physics to modify the Reissner-Nordstrom metric for a charged
black hole\citet{key-20} and derive the Bekenstein-Hawking relation
between entropy and area of a black hole. This is achieved by the
introduction of an additional term, arising from strong interaction,
into the Reissner-Nordstrom metric for a charged black hole which
gives the area of the event horizon of the black hole. For calculation
of the entropy of the black hole, we use the present four dimensional
string theory.

A review of the four dimensional superstring theory will be given
in section-2. In sections-3 and 4, the duality invariant action, coupling
to gravity and generalisation to string theory will be discussed.
In section-5, the equivalence between the two methods will be shown.
In section-6, the metric tensor for a black hole will be obtained
and in section-7, the entropy-area relation for a black hole will
be derived. Section-8 will be devoted to conclusion.

\section{Four dimensional superstring}

We begin with an outline of the method of 4-dimensional string theory
used by us. The Nambu-Goto\citet{key-1,key-2} bosonic string theory,
in the world sheet $\left(\sigma,\tau\right)$ in 26 dimensions, has
the action

\begin{equation}
S_{B}=-\frac{1}{2\pi}\int d^{2}\sigma\left(\partial_{\alpha}X^{\mu}\partial^{\alpha}X_{\mu}\right),\,\,\,\,\,\mu=0,1,\cdots\,,25\label{eq:1}\end{equation}

\begin{flushleft}
where $\partial_{\alpha}=\left(\partial_{\sigma},\partial_{\tau}\right)$.
Using Mandelstam's proof of the equivalence between one boson to two
fermions, in the infinite volume limit in $1+1$ dimensional field
theory, one can write this action as the sum of the action for four
bosonic coordinates $X^{\mu}$ and the action for 44 fermions having
$SO(44)$ symmetry. This is true in finite intervals and circles,
as has beeen shown by Mandelstam\citet{key-5}. The Majorana fermions
can be in the bosonic representation of the Lorentz group $SO(3,1)$.
The 44 fermions form 11 Lorentz vectors. The action can be written
as
\par\end{flushleft}

\begin{equation}
S_{FB}=-\frac{1}{2\pi}\int d^{2}\sigma\,\left[\partial_{\alpha}X^{\mu}\left(\sigma,\tau\right)\partial^{\alpha}X_{\mu}\left(\sigma,\tau\right)-i\sum_{j=1}^{11}\bar{\psi}^{\mu,\, j}\rho^{\alpha}\partial_{\alpha}\psi_{\mu,\, j}\right],\,\,\,\alpha=0,1;\,\,\,\,\mu=0,1,2,3\label{eq:2}\end{equation}

\begin{flushleft}
where $\rho^{0}=\left(\begin{array}{cc}
0 & -i\\
i & 0\end{array}\right),\,\,\,\,\rho^{1}=\left(\begin{array}{cc}
0 & i\\
i & 0\end{array}\right)$ and $\bar{\psi}=\psi^{\dagger}\rho^{0}$. The Dirac operators $\rho^{\alpha}\partial_{\alpha}$
are real since $\rho^{\alpha}$ are imaginary. 
\par\end{flushleft}

\begin{flushleft}
The action given in equation(\ref{eq:2}) is not supersymmetric. In
order to get a supersymmetric action, the eleven $\psi^{\mu,\, j}$
are divided into two species: the six $\psi^{\mu,\, j}$ with $j=1,2,$$\cdots$,
6 and the five $\phi^{\mu,k}\,$ with $k=7,\cdots,11$. For the first
group of six $\psi^{\mu,\, j}=\psi^{\left(+\right)\mu,\, j}+\psi^{(-)\mu,\, j}$
whereas for the second group of five we have $\phi^{\mu,k}=\phi^{(+)\mu,k}-\phi^{(-)\mu,k}$.
Thus they differ in their `neutrino-like' phase. All these are Majorana
fermions. There are $6\times4=24$ `neutrinos' of one type and $5\times4=20$
of the other type. This is indeed possible for the `neutrinos'. The
action can now be written as
\par\end{flushleft}

\begin{equation}
S_{FB}=-\frac{1}{2\pi}\int d^{2}\sigma\,\left[\partial_{\alpha}X^{\mu}\left(\sigma,\tau\right)\partial^{\alpha}X_{\mu}\left(\sigma,\tau\right)-i\sum_{j=1}^{6}\bar{\psi}^{\mu,\, j}\rho^{\alpha}\partial_{\alpha}\psi_{\mu,\, j}+i\sum_{k=7}^{11}\bar{\phi}^{\mu,\, k}\rho^{\alpha}\partial_{\alpha}\phi_{\mu,\, k}\right].\label{eq:3}\end{equation}
which is invariant under $SO(6)\times SO(5)$ as well as $SO(3,1)$.
The action is now supersymmetric and is invariant under the supersymmetric
transformations 

\begin{equation}
\delta X^{\mu}=\bar{\epsilon}\left(e^{j}\psi_{j}^{\mu}-e^{k}\phi_{k}^{\mu}\right),\,\,\,\delta\psi^{\mu,\, j}=-i\epsilon e^{j}\rho^{\alpha}\partial_{\alpha}X^{\mu\,\,\,}\text{and}\,\,\,\delta\phi^{\mu,\, k}=i\epsilon e^{k}\rho^{\alpha}\partial_{\alpha}X^{\mu}.\label{eq:4}\end{equation}
Here $\epsilon$ is a constant anticommuting spinor. The $e^{j}$
are arrays of 11 numbers with ten zeros and only one `1' in the $j$th
place. The $e^{k}$ are arrays of 11 numbers with ten zeros and only
one `$-1$' in the $k$th place. They satisfy the relations $e^{j}e_{j}=6$
and $e^{k}e_{k}=5$. The commutator of two supersymmetric transformations
gives a world sheet transformation. It is to be noted that $\Psi^{\mu}=\left(e^{j}\psi_{j}^{\mu}-e^{k}\phi_{k}^{\mu}\right)$
is the superpartner of $X^{\mu}$. The details can be found in the
references \citet{key-13,key-14,key-15,key-16}. The importance of
the `six' $\psi_{j}^{\mu}$ and the `five' $\phi_{k}^{\mu}$ will
be revealed while deriving the expression for the entropy of a black
hole in section-7. 

The field $X^{\mu}$ can be expressed in terms of the complex coordinates
$z=\sigma+i\tau$ and $\bar{z}=\sigma-i\tau$ as

\begin{equation}
X^{\mu}(z,\bar{z})=x^{\mu}-i\alpha_{0}^{\mu}\ln|z|+i\sum_{m\neq0}\frac{1}{m}\alpha_{m}^{\mu}z^{-m}.\label{eq:4a}\end{equation}

Further,

\begin{equation}
\psi_{\pm}^{\mu,j}(\sigma,\tau)=\frac{1}{\sqrt{2}}\,\,\sum_{m=-\infty}^{\infty}d_{m}^{\mu,j}e^{-im(\sigma\pm\tau)},\,\phi_{\pm}^{\mu,k}(\sigma,\tau)=\frac{1}{\sqrt{2}}\,\,\sum_{m=-\infty}^{\infty}d_{m}^{\prime\mu,k}e^{-im(\sigma\pm\tau)}\,\ldots\,\text{R\,\, sector,}\end{equation}

and

\begin{equation}
\psi_{\pm}^{\mu,j}(\sigma,\tau)=\frac{1}{\sqrt{2}}\,\,\sum_{r\in Z+\frac{1}{2}}b_{r}^{\mu,j}e^{-ir(\sigma\pm\tau)},\,\phi_{\pm}^{\mu,k}(\sigma,\tau)=\frac{1}{\sqrt{2}}\,\,\sum_{r\in Z+\frac{1}{2}}b_{r}^{\prime\mu,k}e^{-ir(\sigma\pm\tau)}\,\ldots\,\text{NS\,\, sector.}\end{equation}

By varying the field and the zweibein, it is seen that the Noether
current $J_{\alpha}$ and the energy momentum tensor $T_{\alpha\beta}$
vanish i.e., 

\begin{equation}
J_{\alpha}=\rho^{\beta}\rho_{\alpha}\Psi^{\mu}\partial_{\beta}X_{\mu}=0,\label{eq:5}\end{equation}
and

\begin{equation}
T_{\alpha\beta}=\partial_{\alpha}X^{\mu}\partial_{\beta}X_{\mu}-\frac{i}{2}\Psi^{\mu}\rho^{\alpha}\partial_{\alpha}\Psi_{\mu}=0.\label{eq:6}\end{equation}
In the light cone coordinates, these become

\begin{equation}
J_{\pm}=\partial_{\pm}X_{\mu}\Psi_{\pm}^{\mu}=0,\label{eq:7}\end{equation}
and 

\begin{equation}
T_{\pm\pm}=\partial_{\pm}X^{\mu}\partial_{\pm}X_{\mu}+\frac{i}{2}\psi_{\pm}^{\mu,\, j}\partial_{\pm}\psi_{\pm\mu,\, j}-\frac{i}{2}\phi_{\pm}^{\mu,\, k}\partial_{\pm}\phi_{\pm\mu,\, k},\label{eq:8}\end{equation}
where $\partial_{\pm}=\frac{1}{2}\left(\partial_{\tau}\pm\partial_{\sigma}\right)$.

The super Virasoro generators of energy momenta $L_{m}$ and the currents
$G_{r},\, F_{m}$ are given by

\begin{eqnarray}
L_{m} & = & \frac{1}{2}\int_{-\pi}^{\pi}d\sigma\, e^{im\sigma}\, T_{++}\nonumber \\
 & = & \frac{1}{2}\sum_{-\infty}^{\infty}:\alpha_{-n}\cdot\alpha_{m+n}:+\frac{1}{2}\sum_{r\in z+\frac{1}{2}}\left(r+\frac{m}{2}\right):\left(b_{-r}\cdot b_{m+r}-b_{-r}^{\prime}\cdot b_{m+r}^{\prime}\right):\,\,\,\,\,\,\,\,\,\text{NS},\label{eq:9}\\
 & = & \frac{1}{2}\sum_{-\infty}^{\infty}:\alpha_{-n}\cdot\alpha_{m+n}:+\frac{1}{2}\sum_{n=-\infty}^{\infty}\left(n+\frac{m}{2}\right):\left(d_{-n}\cdot d_{m+n}-d_{-n}^{\prime}\cdot d_{m+n}^{\prime}\right):\,\,\,\,\text{R,}\label{eq:10}\\
\nonumber \\G_{r} & = & \frac{\sqrt{2}}{\pi}\,\int_{-\pi}^{\pi}d\sigma\, e^{ir\sigma}\, J_{+}=\sum_{-\infty}^{\infty}\alpha_{-n}\left(e^{j}b_{r+n,\, j}-e^{k}b_{r+n,\, k}^{\prime}\right),\,\,\text{NS},\label{eq:11}\\
\text{and}\,\,\,\,\,\,\,\,\,\,\,\,\, F_{m} & = & \sum_{-\infty}^{\infty}\alpha_{-n}\left(e^{j}d_{m+n,\, j}-e^{k}d_{m+n,\, k}^{\prime}\right)\,\,\,\,\,\,\text{\,\,\,\,\,\,\,\,\,\,\,\,\,\,\,\,\,\,\,\,\,\,\,\,\,\,\,\,\,\,\,\,\,\,\,\,\,\,\,\,\,\,\,\,\,\,\,\,\, R.}\label{eq:12}\end{eqnarray}
The normal ordering constant is equal to one.

The physical states $|\phi\rangle$ satisfy the conditions

\begin{equation}
\left(L_{0}-1\right)|\phi\rangle=0,\,\,\,\,\, L_{n}|\phi\rangle=0,\,\,\,\, G_{r}|\phi\rangle=0\,\,\,\,\text{for\,\,\,}n,\, r\,>0,\,\,\,\,\text{NS\,\,\,\, bosonic},\end{equation}

\begin{equation}
L_{n}|\phi\rangle=0,\,\,\,\, F_{n}|\psi\rangle=0,\,\,\,\,\text{for\,\,\,}n>0,\,\,\,\,\text{R\,\,\,\, fermionic.}\end{equation}
So, with $|\psi\rangle=|\psi_{+}\rangle+|\psi_{-}\rangle$, one has

\begin{equation}
\left(F_{0}+1\right)|\psi_{+}\rangle=0\,\,\,\,\,\,\text{and \,\,\,\,\,\,}\left(F_{0}-1\right)|\psi_{-}\rangle=0,\,\,\,\,\,\text{for\,\, R.}\end{equation}
These conditions make the string ghost free. 

The mass spectrum of the model, from the Hamiltonian $L_{0}$ , is
given by

\[
\alpha^{\prime}M^{2}=-1,-\frac{1}{2},0,\frac{1}{2},1,\frac{3}{2},\cdots\,\,\,\,\,\,\,\,\,\,\text{N\, S,}\]
and

\[
\alpha^{\prime}M^{2}=-1,0,1,2,\cdots\,\,\,\,\,\,\,\,\,\,\,\,\,\,\,\,\,\,\,\,\,\,\,\,\,\text{R,}\]
where $\alpha^{\prime}$ is the Regge slope.

The GSO projection eliminates the half integral values and the mass
spectrum is obtained as as $\alpha^{\prime}M^{2}=-1,0,1,2,\cdots$
.

There are, in all, $26\times2=52$ `coordinate like' valued objects.
They are the $2\times4=8$ $\alpha$'s because these are `photon-like',
the $6\times4=24$ $b$ or $d,$ and the $5\times4=20$ $b^{\prime}$
or $d^{\prime}$. So in all there are exactly twice the number of
`objects' in the theory.

The super-Virasora generators described in reference\citet{key-16}
elucidate the method we are following herein.

Before proceeding further, we give a brief account of the preliminaries
of the classical electric and magnetic fields as outlined by Schwarz
and Sen\citet{key-19}.

\section{Duality invariant action and coupling to gravity}

Let us introduce independent gauge fields for the electromagnetic
field and its dual, $A_{\mu}^{(\alpha)}$, $\mu$= 0,1,2,3;~~~~~~~$\alpha=1,2$.
In flat space time the action is

\begin{equation}
S=-\frac{1}{2}\int d^{4}x\,\left(B^{(\alpha)i}\mathcal{L}_{\alpha\beta}E_{i}^{(\beta)}+B^{(\alpha)i}B^{(\alpha)i}\right),\label{eq:16}\end{equation}
where 

\begin{equation}
E_{i}^{(\alpha)}=\partial_{0}A_{i}^{(\alpha)}-\partial_{i}A_{0}^{(\alpha)},\,\,\, B^{(\alpha)i}=\varepsilon^{ijk}\partial_{j}A_{k}^{(\alpha)},\,\,\,\, i,j,k=1,2,3\end{equation}
and \begin{equation}
\mathcal{L}=\left(\begin{array}{cc}
0 & 1\\
-1 & 0\end{array}\right).\end{equation}
Further, \[
\vec{\nabla}\times\vec{B}=\frac{\partial\vec{E}}{\partial t}\,\,\,\,\,\text{and}\vec{\,\,\,\,\,\nabla}\times\vec{E}=-\frac{\partial\vec{B}}{\partial t}.\]

The action given in equation(\ref{eq:16}) is invariant under the
gauge transformation

\begin{equation}
\delta A_{0}^{(\alpha)}=\Psi^{(\alpha)},\,\,\,\text{and\,\,\,\,}\delta A_{i}^{(\alpha)}=\partial_{i}\Lambda^{(\alpha)},\label{eq:19}\end{equation}
where $\Psi^{(\alpha)}$ and $\Lambda^{(\alpha)}$ are the gauge transformation
parameters. We can set

\begin{equation}
A_{0}^{(\alpha)}=0\end{equation}
by using the gauge transformation parameter $\Psi^{(\alpha)}$. By
this choice, we do not lose any equations of motion since $A_{0}^{(\alpha)}$
occurs only as part of a total derivative in the action. The equation
of motion for the field $A_{i}^{(2)}$ is

\begin{equation}
\varepsilon^{ijk}\partial_{j}\left(B^{(2)k}-E_{k}^{(1)}\right)=0.\label{eq:21}\end{equation}
This does not involve any time derivative of $A_{i}^{(2)}$. So, $A_{i}^{(2)}$
can be treated as an auxilary field and be eliminated from the action(\ref{eq:16}).
In order to achieve this we write, from equation(\ref{eq:21}), 

\begin{equation}
B^{(2)k}=E_{k}^{(1)}+\partial_{k}\phi,\label{eq:22}\end{equation}
 for some $\phi$. The $\phi$ in equation(\ref{eq:22}) can be set
to zero by using the freedom associated with the gauge parameter $\Lambda^{(1)}$.
So, we get

\begin{equation}
B^{(2)k}=E_{k}^{(1)}.\label{eq:23}\end{equation}
Putting this in equation(\ref{eq:16}) we get the usual Maxwell action
for the field $A_{\mu}^{(1)},$

\begin{equation}
S_{M}=-\frac{1}{2}\int d^{4}x\,\left(B^{(1)i}B^{(1)i}-E_{i}^{(1)}E_{i}^{(1)}\right),\label{eq:24}\end{equation}
in the gauge $A_{0}^{(1)}=0.$ The duality transformation $\vec{E}\to\vec{B}$
and $\vec{B}\to-\vec{E}$ are manifest and persist in the ongoing
process. 

In order to incorporate gravity, the action(\ref{eq:16}) is generalised
to curved space time in such a way that the $A_{\mu}^{(2)}$ are eliminated
by using their equations of motion and the Maxwell's equations. In
curved space time, the field $A_{\mu}^{(1)}$ is obtained from the
action

\begin{equation}
S_{G}=-\frac{1}{4}\int d^{4}x\,\sqrt{-g}\, g^{\mu\rho}g^{\nu\sigma}F_{\mu\nu}^{(1)}F_{\rho\sigma}^{(1)},\label{eq:25}\end{equation}
where

\begin{equation}
F_{\mu\nu}^{(1)}=\partial_{\mu}A_{\nu}^{(1)}-\partial_{\nu}A_{\mu}^{(1)}.\end{equation}

\section{The string theory action}

Generalisation of the duality invariant action to string theory has
been made mostly by Schwarz, Sen and Maharana\citet{key-17,key-18,key-19}.
Their results, which are specific to our purpose, are summarised below.

The low energy effective action is\citet{key-19}, 

\begin{eqnarray}
S & = & -\frac{1}{2}\int d^{4}x\sqrt{-g}\nonumber \\
 &  & \left[R-\frac{1}{2\left(\ \lambda_{2}\right)^{2}}g^{\mu\nu}\partial_{\mu}\lambda\partial_{\nu}\bar{\lambda}-\frac{1}{4}\lambda_{2}F_{\mu\nu}^{a}\left(LML\right)_{ab}F^{b\mu\nu}+\frac{1}{4}\lambda_{1}F_{\mu\nu}^{a}L_{ab}\tilde{F}^{b\mu\nu}+\frac{1}{8}g^{\mu\nu}\text{Tr}\left(\partial_{\mu}ML\partial_{\nu}ML\right)\right],\label{eq:27}\end{eqnarray}
where $A_{\mu}^{a},\,\, a=1,2,\,\cdots\,,\,12$ are a set of 12 abelian
gauge fields and 

\begin{equation}
F_{\mu\nu}^{a}=\partial_{\mu}A_{\nu}^{a}-\partial_{\nu}A_{\mu}^{a},\,\,\,\tilde{F}^{a\mu\nu}=\frac{1}{2\sqrt{-g}}\varepsilon^{\mu\nu\rho\sigma}F_{\rho\sigma}^{a}.\label{eq:28}\end{equation}
Further,

\begin{equation}
\lambda=\lambda_{1}+i\lambda_{2},\end{equation}
is a complex scalar field, which is like a scalar dilaton field. The
matrix $L$ is given by

\begin{equation}
L=\left(\begin{array}{cc}
0 & I_{6}\\
I_{6} & 0\end{array}\right),\end{equation}
and $M$ is a $12\times12$ matrix valued scalar field satisfying
the constraints

\begin{equation}
M^{T}=M,\,\,\, M^{T}LM=L.\end{equation}
The matrix $M$ can be parametrised as

\begin{equation}
M=\left(\begin{array}{cc}
\hat{G}^{-1} & \hat{G}^{-1}\hat{B}\\
-\hat{B}\hat{G}^{-1} & \hat{G}^{}-\hat{B}\hat{G}^{-1}\hat{B}\end{array}\right),\end{equation}
where $\hat{G}$ are $6\times6$ symmetric matrices and $\hat{B}$
are $6\times6$ antisymmetric matrices. We also have 

\begin{equation}
M^{-1}=\left(\begin{array}{cc}
\hat{G}^{}-\hat{B}\hat{G}^{-1}\hat{B} & -\hat{B}\hat{G}^{-1}\\
\hat{G}^{-1}\hat{B} & \hat{G}^{-1}\end{array}\right)=\eta M\eta,\end{equation}
with $\eta=\left(\begin{array}{cc}
0 & 1\\
1 & 0\end{array}\right)$. The equations of motion obtained from the action(\ref{eq:27}) have
a further $SL(2,R)$ symmetry

\begin{equation}
\lambda\to\frac{a\lambda+b}{c\lambda+d},\,\, F_{\mu\nu}^{a}\to\left(c\lambda_{1}+d\right)F_{\mu\nu}^{a}+c\lambda_{2}\left(ML\right)_{ab}\tilde{F}_{\mu\nu}^{b},\,\,\,\,\,\text{with\,\,\,\,}\, ad-bc=1.\label{eq:34}\end{equation}
If we choose $a=0,\, b=1,\, c=-1\,\,\text{and\,\, }$ $d=0$, the
transformations(\ref{eq:34}) take the form

\begin{equation}
\lambda\to-\frac{1}{\lambda},\,\, F_{\mu\nu}^{a}\to-\lambda_{1}F_{\mu\nu}^{a}-\lambda_{2}\left(ML\right)_{ab}\tilde{F}_{\mu\nu}^{b}.\label{eq:35}\end{equation}
With the choice $\lambda_{1}=0,$ the transformation(\ref{eq:35})
takes the electric field to magnetic field and vice versa. It also
takes $\lambda_{2}\to\frac{1}{\lambda_{2}}$. Since $\frac{1}{\lambda_{2}}=\alpha^{\prime}$
can be identified with the coupling constant of the string theory,
the duality transformation(\ref{eq:35}) takes a strong coupling theory
to a weak coupling theory and vice versa. Thus, equation(\ref{eq:35})
contains the strong-weak coupling duality transformation as well as
the electric-magnetic duality transformation.

The action(\ref{eq:27}) has manifest invariance under SL(2,R) and
global coordinate transformation, but O(6,6) is only a symmetry of
the equations of motion.

The Ricci tensor $R_{\mu\nu}$ and other relations are obtained as

\begin{equation}
R_{\mu\nu}=\frac{1}{4\left(\lambda_{2}\right)^{2}}\left(\partial_{\mu}\overline{\lambda}\partial_{\nu}\lambda+\partial_{\nu}\overline{\lambda}\partial_{\mu}\lambda\right)+2\lambda_{2}F_{\mu\rho}^{a}\left(LML\right)_{ab}F_{\nu}^{b\rho}-\frac{\lambda_{2}}{2}g_{\mu\nu}F_{\rho\sigma}^{a}\left(LML\right)_{ab}F_{}^{b\rho\sigma},\label{eq:36}\end{equation}

\begin{equation}
D_{\mu}\left(-\lambda_{2}\left(ML\right)_{ab}F^{b\mu\nu}+\lambda_{1}\tilde{F}^{a\mu\nu}\right)=0,\label{eq:37}\end{equation}

\begin{equation}
\frac{D^{\mu}D_{\mu}\lambda}{\left(\lambda_{2}\right)^{2}}+\frac{i}{\left(\lambda_{2}\right)^{3}}D^{\mu}\lambda D_{\mu}\lambda-iF_{\mu\nu}^{a}\left(LML\right)_{ab}F_{}^{b\mu\nu}+\tilde{F}_{\mu\nu}^{a}L_{ab}F^{b\mu\nu}=0,\label{eq:38}\end{equation}
and finally

\begin{equation}
D_{\mu}\tilde{F}^{a\mu\nu}=0,\label{eq:39}\end{equation}
is given by the Bianchi identities satisfied by the $F_{\mu\nu}^{(a)}$.
Here $D_{\mu}$ denotes the standard covariant derivative.

When the field $A_{\mu}^{(2)}$ is eliminated by using its equation
of motion, we get a part of the action as

\begin{equation}
-\frac{1}{4}\int d^{4}x\,\left(\lambda_{2}F_{\mu\nu}^{(1)}F_{\rho\sigma}^{(1)}-\lambda_{1}F_{\mu\nu}^{(1)}\tilde{F}_{\rho\sigma}^{(1)}\right)\eta^{\mu\rho}\eta^{\nu\sigma}.\label{eq:40}\end{equation}
This is the gauge field dependent part of the action(\ref{eq:27})
in flat background. The gauge field $A_{\mu}^{(2)}$ will be identified
with the dual vector potential, introduced by Kallosh and Ortin\citet{key-21}.

For $M=I,\, L=I,$ there are precisely 22 copies of this action along
with the four copies due to the $\lambda$'s. It may be pointed out
again that

\begin{equation}
B^{(a,\alpha)\, i}=\varepsilon^{ijk}\partial_{j}A_{k}^{(a,\alpha)},\,\,\, E_{i}^{(a,\alpha)}=\partial_{0}A_{i}^{(a,\alpha)}-\partial_{i}A_{0}^{(a,\alpha)}.\end{equation}

\section{Equivalence of the two methods}

\textit{The string states and quantum operators}:

We give a brief accounrt of the correspondence between the string
states and quantum operators. It is known that the Neveu-Schwarz bosonic
sector, in the bosonic superstring, contains a bosonic tachyon. The
ground state of zero mass can be constructed conveniently by using
this tachyon state. The vacuum state $|0,0\rangle$ of the string
is the functional integral of the string theory over a semi-infinite
strip, which can be conformally mapped to the unit circle. The following
recipe, provided by Polchinski\citet{key-22}, for the link between
superstring states and quantum operators is very useful for quantising
our theory.

Radial quantisation has a natural isomorphism between the string state
space of Conformal Field Theory (CFT) in a periodic spatial dimension
and the space of local operators. Let a local isolated operator $\mathcal{A}$
be considered at the origin of the unit circle $|z|=1$, with no more
inside and with no other specification outside the circle. Let us
open a slit in the circle and consider the path integral, on the unit
circle, giving an inner product $\langle\psi_{out}|\psi_{in}\rangle.$
Here, $\psi_{in}$ is the incoming state given by the path integral
$|z|\,<1$ and $\psi_{out}$ is the outgoing state at $|z|>1.$ Thus
we see that a field $\phi$ is decomposed into integrals outside,
inside and on the circle. We denote the last one by $\phi_{B}.$ The
outside and inside integrals are denoted by $\psi_{out}(\phi_{B})$
and $\psi_{in}(\phi_{B})$ respectively. The remainder is $\int\left[d\phi_{B}\,\right]\psi_{out}(\phi_{B})\psi_{in}(\phi_{B}).$
The incoming state is denoted by $|\ \psi_{A}\rangle$ since it depends
on the operator $\mathcal{A}.$ This gives the required mapping from
operators to states. Summarizing,`` the mapping from operators to
states is given by a path integral on the unit disk''. The inverse
mapping is also true. 

For any conserved charge $Q$, the operator equivalent of ~$Q\mathcal{A}$
is $Q|\psi_{A}\rangle.$ For example, if $\mathcal{A}$ is the unit
operator $\hat{1}$, and $Q=\alpha_{m}=\oint dz\,(2\pi)^{-1}z^{m}\partial X$
~for $m\geq0,$ so that $\partial X$ is analytic and the integral
vanishes for $m\geq0$, we get $\alpha_{m}|\psi_{\hat{1}}\rangle=0,$
for $m\geq0.$ The exact correspondence between the unit operator
$\hat{1}$ and the string vacuum $|0,0\rangle$ is thus established.

\begin{equation}
\hat{1}\leftrightarrow|0,0\rangle\end{equation}

Similarly, the operator equivalence of the state $|0,k\rangle$ is
given by

\begin{equation}
:e^{ik\cdot X(z)}:\leftrightarrow|0,k\rangle.\end{equation}
where $X(z)$ is given by equation(\ref{eq:4a}). The expression $:e^{ikX}:$
implies normal ordering of the operators contained in it. In the state
$|0,k\rangle$, the first symbol refers to the value of $m$ and the
second one to the eigenvalue of $\alpha_{0}^{\mu}$, i.e., $\alpha_{0}^{\mu}|0,k\rangle=k^{\mu}|0,k\rangle.$
So, for the tachyon, $|0,k\rangle\leftrightarrow e^{ikx}$ since $|z|=1$
on the circumference of the circle, the tachyonic vacuum cannot annihilate\citet{key11}.
The CFT unitarity gives the normalization

\begin{equation}
\langle0,k|0,k^{\prime}\rangle=2\pi\delta(k-k^{\prime})\end{equation}
For the three spatial components one has

\begin{equation}
\langle0,\vec{k}|0,\vec{k^{\prime}}\rangle=(2\pi)^{3}\delta^{(3)}(k-k^{\prime})\end{equation}
This is generalized to the normalization of massless states with $k_{0}=|\vec{k}|$,
and we use one like the normalization for massive vector meson, i.e.,

\begin{equation}
\langle0,\vec{k}|0,\vec{k^{\prime}}\rangle=(2\pi)^{3}\left(2k_{0}\right)\delta^{(3)}(k-k^{\prime}).\end{equation}
\textit{The equivalence of the two methods}:

The equivalence between the two methods is summarised below.

We have $(12+1)\times4=52$ objects whose equations of motion are
written down in both the theories, and they are equivalent in the
sense that the quantities we wish to determine would be meaningfully
the same in both the methods. This implies the following way of expressing
the $F_{\mu\nu}^{a}$ 's.

\begin{alignat}{1}
F_{\mu\nu}^{(1)}(x) & =\int\frac{d^{4}k}{\left(2\pi^{4}\right)}\,\left(k_{\mu}\alpha_{\nu}-k_{\nu}\alpha_{\mu}\right)\,|0,k\rangle e^{ikx}=\left(\partial_{\mu}A_{\nu}^{(1)}-\partial_{\nu}A_{\mu}^{(1)}\right),\label{eq:42}\\
F_{\mu\nu}^{(j)}(x) & =\int\frac{d^{4}k}{\left(2\pi^{4}\right)}\,\left(k_{\mu}b_{\nu}^{j}-k_{\nu}b_{\mu}^{j}\right)\,|0,k\rangle e^{ikx}=\left(\partial_{\mu}A_{\nu}^{(j)}-\partial_{\nu}A_{\mu}^{(j)}\right),\, j=2,\,\cdots\,,7...(NS),\label{eq:43}\\
F_{\rho\sigma}^{(j)}(x) & =\int\frac{d^{4}k}{\left(2\pi^{4}\right)}\,\left(k_{\rho}d_{\sigma}^{j}-k_{\sigma}d_{\rho}^{j}\right)\,|0,k\rangle e^{ikx}=\left(\partial_{\rho}A_{\sigma}^{(j)}-\partial_{\sigma}A_{\rho}^{(j)}\right),\, j=2,\,\cdots\,,7....(R),\label{eq:44}\\
F_{\mu\nu}^{(r)}(x) & =\int\frac{d^{4}k}{\left(2\pi^{4}\right)}\,\left(k_{\mu}b_{\nu}^{\prime r}-k_{\nu}b_{\mu}^{\prime r}\right)\,|0,k\rangle e^{ikx}=\left(\partial_{\mu}A_{\nu}^{(r)}-\partial_{\nu}A_{\mu}^{(r)}\right),\, r=8,\,\cdots\,,12...(NS),\label{eq:45}\\
F_{\rho\sigma}^{(r)}(x) & =\int\frac{d^{4}k}{\left(2\pi^{4}\right)}\,\left(k_{\rho}d_{\sigma}^{\prime r}-k_{\sigma}d_{\rho}^{\prime r}\right)\,|0,k\rangle e^{ikx}=\left(\partial_{\rho}A_{\sigma}^{(r)}-\partial_{\sigma}A_{\rho}^{(r)}\right),\, r=8,\,\cdots\,,12....(R).\label{eq:46}\end{alignat}
They are suitably rearranged in the theory given in section-3. The
extra bosons $\left(\eta,\bar{\eta}^{\prime}\right)$ are computed
separately. It should be noted that the mapping from operators to
states is given by path integrals on the unit disc as described by
Polchinski\citet{key-22}. The inverse is also true. The tachyonic
states become very useful to construct the ground state of zero mass.
Thus the correspondence and equivalence between the two methods are
easily established. The $F_{\mu\nu}^{(a)}(x)$ 's given in equations(\ref{eq:42})
to (\ref{eq:46}) are equivalent to the 12 fields $F_{\mu\nu}^{a}$
's given in equation(\ref{eq:28}). So our results for these 12 fields
are the same as that of the 4 dimensional theory. Further, $\lambda_{1}$
and $\lambda_{2}$ have the same expressions as used by us previously.
One is related to the dilaton field and the other to the basic couplng
constant $\lambda_{2}=\frac{1}{\alpha^{\prime}}$.

It is an important fact to realise that a way has been found out to
understand how duality is brought into the picture. We would like
to examine how the strong-weak coupling duality or the electric-magnetic
duality goes through to the very basis of the string theory. Further,
the general theory of relativity has been introduced very conveniently
and one can proceed to study the gravitational effects more closely
and effectively. Inclusion of gravity and the strong-weak coupling
duality are the two important attributes of the four dimensional string
theory presented here.

\section{The metric tensor for black hole}

Here we show how the metric tensor for a charged black hole (the Reissner-Nordstrom
metric) gets modified due to the incorporation of strong interaction
effects. The metric tensor, in general, is given by

\begin{equation}
d\tau^{2}=g_{\mu\nu}dx^{\mu}dx^{\nu},\label{eq:47}\end{equation}
with $x^{\mu}=(t,r,\theta,\phi)$ . The spherically symmetric, static
metric is written in a convenient form as

\begin{equation}
d\tau^{2}=-e^{2\nu}dt^{2}+e^{2\mu}dr^{2}+r^{2}\left(d\theta^{2}+\sin^{2}\theta\, d\phi^{2}\right).\label{eq:48}\end{equation}
where $\mu$ and $\nu$ are functions of $r,t$ . For static, spherically
symmetric case, the vanishing of the Ricci tensor $R_{22}=0$ in the
field equation shows that both $\mu$ and $\nu$ are independent of
time and depend only on $r$. Further, from the relation $R_{00}+R_{11}=0$
for the Ricci tensor, we get $\mu=-\nu$. This leads to the Schwarzschild
metric

\begin{equation}
d\tau^{2}=-\left(1-\frac{2M}{r}\right)dt^{2}+\left(1-\frac{2M}{r}\right)^{-1}dr^{2}+r^{2}\left(d\theta^{2}+\sin^{2}\theta\, d\phi^{2}\right)\end{equation}
where $M$ is the mass located at the origin.

This metric gets modified for a charged black hole. For a static,
spherically symmetric charge distribution with total charge $Q$,
the only non-vanishing component of the field strength tensor is $F_{01}=-\frac{Q}{r^{2}}$.
So, in the Einstein field equation, instead of $R_{22}=0$, we should
have $R_{22}=\left(F_{01}\right)^{2}=\frac{Q^{2}}{r^{4}}.$ This gives
the usual result\citet{key-20} for the Ricci tensor

\begin{equation}
R_{22}=\frac{1}{r^{2}}\left(1-e^{2\nu}\right)-\frac{2}{\nu}e^{2\nu}\frac{d\nu}{dr},\label{eq:49}\end{equation}
so that

\begin{equation}
e^{2\nu}=1-\frac{2M}{r}-\frac{1}{r}\int_{\infty}^{r}r^{\prime2}R_{22}(r^{\prime})dr^{\prime}=1-\frac{2M}{r}+\frac{Q^{2}}{r^{2}}=e^{-2\mu}.\label{eq:50}\end{equation}
Thus the metric for a charged black hole,with total charge $Q$, and
mass $M$ is given by the Reissner-Nordstrom metric

\begin{equation}
d\tau^{2}=-\left(1-\frac{2M}{r}+\frac{Q^{2}}{r^{2}}\right)dt^{2}+\left(1-\frac{2M}{r}+\frac{Q^{2}}{r^{2}}\right)^{-1}dr^{2}+r^{2}\left(d\theta^{2}+\sin^{2}\theta d\phi^{2}\right).\label{eq:51}\end{equation}
In addition to this, we should have a contribution from the strong
interaction. For higher energy systems like black hole\citet{key-18},
$F_{0r}\sim\frac{1}{r^{3}}\,\,\,\text{and\,\,\,}\widetilde{F}_{0r}\sim\frac{1}{r^{2}}$
so that a term containing the factor $\frac{1}{r^{5}}$ is also present
in $R_{22}.$ 

For a static, constant background field, the field strength tensor
is \citet{key-23}

\begin{equation}
F_{\alpha\mu\nu}=C_{\alpha\beta\gamma}A_{\beta\mu}A_{\gamma\nu,}\end{equation}
with gauge covariant derivative 

\begin{equation}
D_{\lambda}F_{\delta\mu\nu}=C_{\delta\epsilon\gamma}C_{\alpha\beta\gamma}A_{\epsilon\nu}A_{\beta\mu}A_{\lambda\alpha}.\end{equation}
Here, $C_{\alpha\beta\gamma}$ are the structure constants. From perturbation
expansion it is clear that the coefficient of ~~$A_{\alpha\beta}A_{\beta\gamma}A_{\gamma\mu}$
is

\begin{equation}
C_{\alpha\beta\gamma}C_{\beta\gamma\mu}\sim N\,\delta_{\alpha\mu,}\label{eq:54}\end{equation}
where $N=\alpha^{\prime}M^{2}=0,\,1,\,2,\,\cdots$ is an integer.
Here $\alpha^{\prime}=\frac{1}{2}$ is the Regge slope . It is assumed
that the mass $M$ of the black hole falls on the Regge trajectory
or is very close to it. Thus in the presence of both electromagnetic
and strong interaction the Ricci tensor $R_{22}$ is given by

\begin{equation}
R_{22}=\frac{Q^{2}}{r^{4}}-2\frac{\lambda_{\text{v}}}{\alpha^{\prime}}\frac{N}{r^{5}}.\end{equation}
The constant $\lambda_{\text{v}}$, in the above equation, is determined
from the condition that the black hole is to be extremal. Equation(\ref{eq:50})
now becomes

\begin{equation}
e^{2\nu}=1-\frac{2M}{r}-\frac{1}{r}\int_{\infty}^{r}r^{\prime2}R_{22}(r^{\prime})dr^{\prime}=1-\frac{2M}{r}-\frac{1}{r}\int_{\infty}^{r}r^{\prime2}\left(\frac{Q^{2}}{r^{\prime4}}-2\frac{\lambda_{\text{v}}}{\alpha^{\prime}}\frac{N}{r^{\prime5}}\right)dr^{\prime}=1-\frac{2M}{r}+\frac{Q^{2}}{r^{2}}-\frac{\lambda_{\text{v}}N}{\alpha^{\prime}}\frac{1}{r^{3}}\label{eq:55}\end{equation}

Due to the inclusion of the effect of strong interaction, the Reissner-Nordstrom
metric is now modified and is given by\citet{key-20}

\begin{equation}
d\tau^{2}=-\left(1-\frac{2M}{r}+\frac{Q^{2}}{r^{2}}-\frac{\lambda_{\text{v}}}{\alpha^{\prime}}\frac{N}{r^{3}}\right)dt^{2}+\left(1-\frac{2M}{r}+\frac{Q^{2}}{r^{2}}-\frac{\lambda_{\text{v}}}{\alpha^{\prime}}\frac{N}{r^{3}}\right)^{-1}dr^{2}+r^{2}\left(d\theta^{2}+\sin^{2}\theta\, d\phi^{2}\right)\label{eq:55a}\end{equation}
We further simplify this as follows. The cubic equation

\begin{equation}
1-\frac{2M}{r}+\frac{Q^{2}}{r^{2}}-\lambda_{\text{v}}\frac{N}{\alpha^{\prime}}\frac{1}{r^{3}}=0\end{equation}
has three roots $r_{0},\, r_{1}$ and $r_{2}$ which satisfy the relations

\begin{equation}
r_{0}+r_{1}+r_{2}=2M,\,\,\, r_{1}r_{2}+r_{0}r_{1}+r_{0}r_{2}=Q^{2}\,\,\text{and\,\,}r_{0}r_{1}r_{2}=\lambda_{\text{v}}\frac{N}{\alpha^{\prime}}.\end{equation}
This leads to the solutions

\begin{eqnarray}
r_{0} & = & \frac{2}{3}M-\frac{1}{6}\sqrt{4M^{2}-3Q^{2}-\frac{3}{2}\frac{N}{\alpha^{\prime}}}+\frac{1}{2}\sqrt{4M^{2}-3Q^{2}+\frac{1}{2}\frac{N}{\alpha^{\prime}}},\\
r_{1} & = & \frac{2}{3}M+\frac{1}{3}\sqrt{4M^{2}-3Q^{2}-\frac{3}{2}\frac{N}{\alpha^{\prime}}},\\
\text{and}\,\,\,\,\,\,\,\,\,\,\,\,\,\,\,\,\,\,\,\,\,\,\,\,\,\,\,\, r_{2} & = & \frac{2}{3}M-\frac{1}{6}\sqrt{4M^{2}-3Q^{2}-\frac{3}{2}\frac{N}{\alpha^{\prime}}}-\frac{1}{2}\sqrt{4M^{2}-3Q^{2}+\frac{1}{2}\frac{N}{\alpha^{\prime}}}.\end{eqnarray}
If the above roots are to be real, one should have $4M^{2}\geq\left(3Q^{2}+\frac{3}{2}\frac{N}{\alpha^{\prime}}\right)$.
In order to calculate the area of the event horizon of the black hole,
one must have at least 

\begin{equation}
4M^{2}=3Q^{2}+\frac{3}{2}\frac{N}{\alpha^{\prime}}.\end{equation}
Further, for an extremal black hole one has $Q=M$ so that 

\begin{equation}
Q^{2}=\frac{3}{2}\frac{N}{\alpha^{\prime}}\end{equation}
Thus, for an extremal black hole, the roots $r_{0},\, r_{1}$ and
$r_{2}$ become

\begin{equation}
r_{0}=\frac{2}{3}M+\sqrt{\frac{N}{2\alpha^{\prime}}},\,\, r_{1}=\frac{2}{3}M\,\,\,\text{and\,\,\,}r_{2}=\frac{2}{3}M-\sqrt{\frac{N}{2\alpha^{\prime}}}.\label{eq:55b}\end{equation}
Substitution of this in the relation $r_{0}r_{1}r_{2}=\lambda_{\text{v}}\frac{N}{\alpha^{\prime}}$,
leads to the value of $\lambda_{\text{v}}$ as

\begin{equation}
\lambda_{\text{v}}=\frac{M}{9}.\end{equation}
So, the metric of equation(\ref{eq:55a}) now becomes 

\begin{equation}
d\tau^{2}=-\left(1-\frac{2M}{r}+\frac{Q^{2}}{r^{2}}-\frac{M}{9\alpha^{\prime}}\frac{N}{r^{3}}\right)dt^{2}+\left(1-\frac{2M}{r}+\frac{Q^{2}}{r^{2}}-\frac{M}{9\alpha^{\prime}}\frac{N}{r^{3}}\right)^{-1}dr^{2}+r^{2}\left(d\theta^{2}+\sin^{2}\theta\, d\phi^{2}\right).\label{eq:56}\end{equation}
 In terms of the solutions $r_{0,\,\,}r_{1\,\,\,}\text{and}\,\,\, r_{2}$
this metric can be recast as

\begin{equation}
d\tau^{2}=-\chi^{-\frac{1}{2}}(r)\left(1-\frac{r_{0}}{r}\right)dt^{2}+\chi^{\frac{1}{2}}(r)\left[\left(1-\frac{r_{0}}{r}\right)^{-1}dr^{2}+\chi^{-\frac{1}{2}}(r)r^{2}\left(d\theta^{2}+\sin^{2}\theta\, d\phi^{2}\right)\right]\end{equation}
where $\chi^{-\frac{1}{2}}(r)=\left(1-\frac{r_{1}}{r}\right)\left(1-\frac{r_{2}}{r}\right).$
So, the metric coefficients are

\begin{equation}
g_{\theta\theta}\rightarrow r^{2}\left(1-\frac{r_{1}}{r}\right)\left(1-\frac{r_{2}}{r}\right),\,\,\,\,\, g_{\phi\phi}\rightarrow r^{2}\left(1-\frac{r_{1}}{r}\right)\left(1-\frac{r_{2}}{r}\right)\sin^{2}\theta.\end{equation}
The area of the event horizon of the black hole is

\begin{equation}
A=\int\sqrt{g_{\theta\theta\,}g_{\phi\phi}}\mid_{r=r_{0}}d\theta\, d\phi=4\pi\left(r_{0}-r_{1}\right)\left(r_{0}-r_{2}\right)\end{equation}
which, by using equation(\ref{eq:55b}) comes out to be

\begin{equation}
A^{open}=4\pi\frac{N}{\alpha^{\prime}}=4\pi M\sqrt{\frac{N}{\alpha^{\prime}}}=4\pi M\sqrt{2N}.\label{eq:57}\end{equation}
In this case, the horizon is visible\citet{key-24}.

For closed string, we have $\alpha^{\prime}M^{2}=2\left(N_{L}+N_{R}\right)$
instead of $\alpha^{\prime}M^{2}=N.$ So the area of the event horizon,
for closed string, becomes

\begin{equation}
A^{close}=8\pi M\left(\sqrt{N_{L}}+\sqrt{N_{R}}\right).\label{eq:58}\end{equation}
We shall see, in the next section, that the area of the event horizon
of the black hole, in string theory,  is related to the black hole
entropy.

\section{Black hole entropy}

In order to calculate the entropy of the black hole, it is necessary
to enumerate the physical modes of the string and one has to use the
26 dimensional theory. There are 24 physical bosons in the 26-D Nambu-Goto
bosonic string. Since the total normal ordering constant has the value
$a=-1$, the normal ordering constant for each boson is equal to $-\frac{1}{24}$.
The total number of open string bosonic states $d_{n}$ can be obtained
from the generating function\citet{key-12}

\begin{equation}
G(\omega)=\sum_{n=0}^{\infty}d_{n}\omega^{n}=\text{tr}\,\omega^{N},\label{eq:59}\end{equation}
which, in turn, is evaluated from the following.

\begin{equation}
\text{Tr\,}\omega^{N}=\prod_{n=1}^{\infty}\left(1-\omega\right)^{-N}=\left(f\,(\omega)\right)^{-N}=\left(f\,(\omega)\right)^{-24},\end{equation}
where $f(\omega)=\prod_{n=1}^{\infty}\left(1-\omega^{n}\right)$ is
the classical partition function. The number of states $d_{n}$ can
be projected out from $G(\omega)=\sum_{n=0}^{\infty}d_{n}\omega^{n}$
by a contour integral along a small circle about the origin,

\begin{equation}
d_{n}=\frac{1}{2\pi i}\oint\frac{G(\omega)}{\omega^{n+1}}\, d\omega.\label{eq:61}\end{equation}
One finds that for $n\to\infty$ 

\begin{equation}
d_{N}\sim e^{\pi\sqrt{2N}}.\end{equation}
So, in case of open string, the black hole entropy is

\begin{equation}
S^{open}=M\,\ln\, d_{N}\,=\pi M\sqrt{2N}.\label{eq:63}\end{equation}
Fron equations(\ref{eq:57}) and (\ref{eq:63}) we get the relation
between the entropy and area of a black hole as

\begin{equation}
S^{open}=\frac{A^{open}}{4},\label{eq:64}\end{equation}
which is the correct Bekenstein-Hawking relation between entropy and
area of a black hole.

This result was obtained in 26 dimensions where we have used 26 bosonic
coordinates. Now we proceed to evaluate the entropy of black hole
using the 4 dimensional superstring theory which has 4 bosonic modes
and 4 fermionic modes\citet{key-20}. The degeneracy $d_{n}$ is obtained
from the generating function

\begin{equation}
G(\omega)=\sum_{n=0}^{\infty}d_{n}\omega^{n}=\text{tr}\,\omega^{N}=4\prod_{N=1}^{\infty}\left(\frac{1+\omega^{N}}{1-\omega^{N}}\right)^{4}\label{eq:65}\end{equation}
Asymptotically, i.e., as $\omega\to1$, we have

\begin{equation}
G(\omega)\sim e^{2\pi^{2}/(1-\omega)},\end{equation}
which yields \begin{equation}
d_{N}=\frac{1}{2\pi i}\oint\frac{G(\omega)}{\omega^{N}+1}d\omega\sim e^{\pi\sqrt{2N}}.\end{equation}

For a closed string, we have $\alpha^{\prime}M^{2}=2\left(N_{L}+N_{R}\right)$
instead of $\alpha^{\prime}M^{2}=N$. In this case, the level density,
again being statistical, is given by 

\begin{equation}
d_{N}^{\text{close}}=d_{N_{L}}\cdot d_{N_{R}}\sim\exp\left(2\pi\left(\sqrt{N_{L}}+\sqrt{N_{R}}\right)\right).\label{eq:68}\end{equation}
The corresponding entropy is

\begin{equation}
S^{\text{close}}=M\,\ln\, d_{N}^{\text{close}}=2\pi M\left(\sqrt{N_{L}}+\sqrt{N_{R}}\right).\label{eq:69}\end{equation}
From equations(\ref{eq:69}) and (\ref{eq:58}), we get the entropy-area
relation for extremal black hole, for closed string, as

\begin{equation}
S^{\text{close}}=\frac{A^{\text{close}}}{4}.\label{eq:70}\end{equation}

This result is exactly the same as that given in equation(\ref{eq:64})
for open string. Thus the equations(\ref{eq:70}) and (\ref{eq:64})
give the correct Bekenstein-Hawking relation between entropy and area
of a black hole.

\section{Conclusion}

The extended string theory is thus seen to yield the expected results
and should be pursued vigorously. The original pure bosonic string
theory can be turned into a superstring. Again, without much ado,
we can go over to a four dimensional string theory with gravity. The
full consequences of the theory can be realised and we can get the
correct metric tensor as well as the entropy-area relation for a charged
extremal black hole. 

For $D=4$, as in this case, there is the $SU(1,1)$ or $SL(2,R)$
symmetry. The `dilaton' and the `axion' are present together and magically
parametrize the coset space as has been stated by Maharana and Schwarz\citet{key-17}.

We believe that the theory presented here is true to all orders in
perturbation theory and not for a limited range, as has been stressed
only for the ten dimensional theories. In four dimensions, both the
approaches are complimentary and should be taken equally seriously.

\end{document}